\documentclass[a4paper]{article}
\usepackage{amsthm}
\usepackage{amssymb}
\usepackage[small]{titlesec}
\usepackage{chngcntr}
\counterwithin*{equation}{section}

\usepackage[colorlinks]{hyperref}
\hypersetup{linkcolor={blue}}

\usepackage{graphicx}

\newcommand{\R}{\mathbb{R}}

\def \A{{\rm A}}
\def \B{{\rm B}}
\def \O{{\rm O}}

\def \I{{\cal I}}
\def \d{{\rm d}}

\title{\large{\bf Lambert's theorem and projective dynamics}}

\author{\normalsize{\bf Alain Albouy$^1$ and Lei Zhao$^2$}\\
\\
\normalsize{$^1$ IMCCE, UMR 8028,}\\
\normalsize{77, avenue Denfert-Rochereau}\\
\normalsize{F-75014 Paris}\\
\normalsize{Alain.Albouy@obspm.fr}\\
\\
\normalsize{$^2$ Institute of Mathematics}\\
\normalsize{University of Augsburg}\\
\normalsize{Augsburg, Germany}\\
\normalsize{Lei.Zhao@math.un-augsburg.de}}

\date{}

\begin{document}

\maketitle

\section*{Abstract}  We prove that the classical Lambert theorem about the elapsed time on an arc of Keplerian orbit extends without change to the Kepler problem on a space of constant curvature. We prove that the Hooke problem has
a property similar to Lambert's theorem, which also extends to the spaces of constant curvature.

\section{Introduction}

Johann Heinrich Lambert (1728-1777) is among the famous authors who improved our understanding of Euclid's fifth postulate. He wrote in 1766 the essay {\it Die Theorie der Parallellinien}, which was published posthumously in 1786.

Lambert is also well known as the discoverer, in 1761, of a strange and useful property of the Keplerian motion [1]. We will prove that this property is still valid without the fifth postulate, i.e.\ for the Keplerian motion on a space of constant curvature. This generalized Kepler problem was introduced in 1859 by Paul Serret [1]. Since that time, it has drawn the attention of mathematicians and physicists due to its numerous analogies with the usual Kepler problem in the Euclidean space. This quest for analogies between the flat case and the curved case was extended successfully to other systems of classical and quantum mechanics.  Clifford [1] introduced the free motion of a rigid body on the sphere (see Borisov \& Mamaev [1]). Killing [1] introduced and solved the two fixed centre problem in the constant curvature spaces (see Borisov, Mamaev \& Bizyaev [1]).  Schr\"odinger [1] found the spectrum of  the hydrogen atom after this warning: 

\small

\begin{quote}

It may appear foolish to pay attention to the extremely feeble curvature of the Universe in dealing with the hydrogen atom, because even the influence of those much stronger fields of gravitation in which all our observations are actually made is (if the frame is properly chosen) entirely negligible. But this problem, by obliterating the sharp cut between ``elliptic and hyperbolic orbits'' (the classical orbits here are {\it all} closed) and by resolving the continuous spectrum into an intensely crowded line spectrum, has extremely interesting features, well worth investigating by a method which proves hardly more complicated here than in the flat case. 

\end{quote}

\normalsize

As far as we know, Lambert's property was never considered in this perspective. In the usual Kepler problem, this property is known as Lambert's theorem. We will also consider the {\it Hooke problem}, i.e.\ the motion of a particle in the plane attracted by a force vector proportional to the position vector. We will show that  this problem and its curved analogues have a similar property, which we will define in a general context and call the {\it Lambert-Hamilton property}, thus acknowledging an important remark due to Hamilton in 1834.

As a kind of explanation of these results, we will show the good compatibility of some classical {\it transformations} with the Lambert-Hamilton property. In this work, we consider the projective transformations. Their striking properties appeared in a remark by Halphen [1] about the Kepler problem and the Hooke problem in 1878. Appell [1] gave a general framework in 1890. We will show that if Appell's projection sends a natural mechanical system with the Lambert-Hamilton property onto another natural mechanical system, then this new system also has the Lambert-Hamilton property.

In a forthcoming work, we will consider the conformal transformations. The transformation  of the complex plane $z\mapsto z^2$ sends the Hooke problem onto the Kepler problem, as discovered by Maclaurin [1] in 1742. It has been commonly used as a regularizing transformation since Levi-Civita [1]. We will explain in a general framework, proposed by Goursat  [1] and Darboux [1] in 1889, why this kind of transformation respects the Lambert-Hamilton property even better than the projective transformations. In the present work, we do not relate the Kepler problem and the Hooke problem, but rather prove by a short computation that the Hooke problem possesses this property, and use the projective transformations to pass to the curved cases.

\subsection{What is the Lambert-Hamilton property?}

J.H.\ Lambert discovered and proved in 1761 the following strange property of the Keplerian motion around a centre $\O$. We denote by $d(\A,\B)$ the distance between two points $\A$ and $\B$ in the Euclidean plane.

\medskip
{\bf Theorem 1.1 (Lambert).} The time required to reach a point $\B$
from a point $\A$ with a given total energy $H$, under the Newtonian attraction of a fixed centre $\O$, does not vary if we change continuously $\A$ and $\B$ while keeping the distances $d(\A,\B)$ and  $d(\O,\A)+d(\O,\B)$ constant.

\medskip
{\bf Remark 1.2.} We should insist that the above allowed changes of $\A$ and $\B$ do not depend on the value of the energy $H$, since they are only subjected to the two conditions that $d(\A,\B)$ and  $d(\O,\A)+d(\O,\B)$ remain constant. They are the same for any arc of orbit going from $\A$ to $\B$.  Such an arc changes while $\A$ and $\B$ are changing. The change of arc is only subjected to two conditions: $\A$ and $\B$ are its ends, and the energy $H$ does not change. An example of a continuous change of an arc is illustrated by Figure~1.

\vspace{0.5cm}
\centerline{\includegraphics[width=60mm]{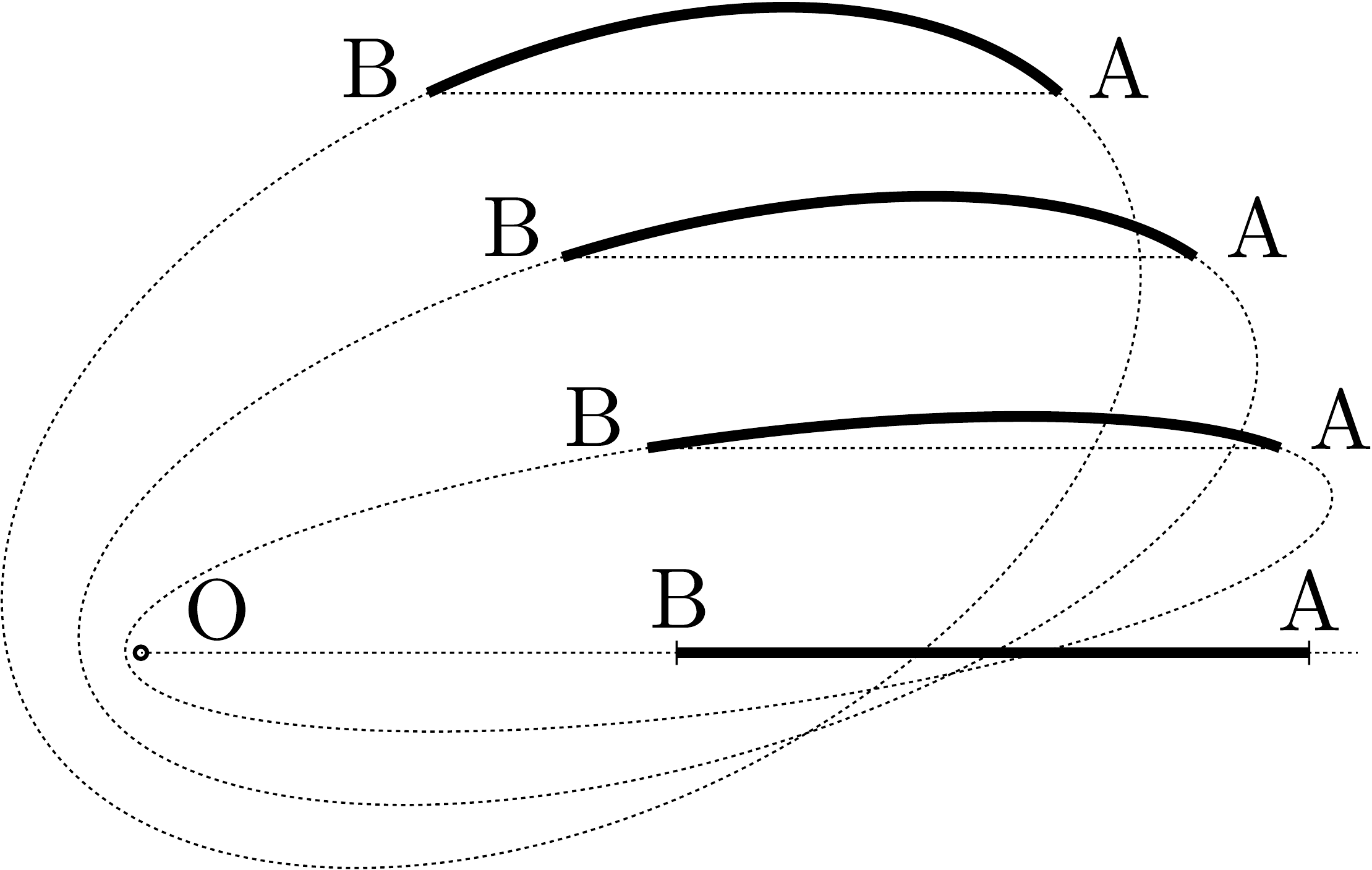}}
\centerline{\it Figure 1. Four Keplerian arcs with same $d(\A,\B)$, $d(\O,\A)+d(\O,\B)$ and $H$.}
\vspace{0.5cm}

{\bf Remark 1.3.} A classical way to state Lambert's theorem is: {\it The time required to reach a point $\B$ from a point $\A$ under the Newtonian attraction of a fixed centre $\O$ only depends on the distances $d(\A,\B)$ and $d(\O,\A)+d(\O,\B)$ and on the total energy $H$.} However, fixing the latter three quantities determines several disconnected components in the space of Keplerian arcs, on which this time is not the same. This is why we state the theorem in terms of continuous changes. See Albouy [3] for more information.

\medskip
{\bf Notation.} We denote by $v=\dot q$ the velocity of a body $q$, and $r=\|q\|$ its distance to the origin $\O$. The energy (or total energy) in a Keplerian motion is $H(q,\dot q)=\|\dot q\|^2/2-1/r$, and the {\it Lagrangian} is $L(q,\dot q)=\|\dot q\|^2/2+1/r$.

In 1834, Hamilton [1] discovered another property of the same changes of arc. Two other quantities, called the action integrals, are also constant under the hypothesis of Theorem~1.1. 

\medskip
{\bf Theorem 1.4 (Hamilton).} The Lagrangian action $S$ and the Maupertuis action $w$, respectively defined as the integrals
\begin{equation}
S=\int_{t_\A}^{t_\B}L\,\d t,\qquad w=\int_{t_\A}^{t_\B}\|\dot q\|^2 \d t,\end{equation}
computed from a point $\A$
to a point $\B$ along a Keplerian arc around $\O$ with a given total energy $H$, do not vary if we continuously change $\A$ and $\B$ while keeping $d(\A,\B)$ and  $d(\O,\A)+d(\O,\B)$ constant.

\medskip
We will generalize the previous theorems to certain other systems, guided by an observation on the classical {\it variation formula}, emphasized by Hamilton in the same article [1] of 1834, and in its continuation [2] of 1835:
\begin{equation}\label{deltaS}
\delta S=\langle \delta \B,v_\B\rangle-\langle \delta \A, v_\A\rangle-H(\delta t_\B-\delta t_\A).
\end{equation}
This observation is the following: in the Kepler problem, the ``infinitesimal changes'' $\delta \A$ and $\delta \B$ of the ends $\A$ and $\B$, allowed by Lambert's conditions in his theorem, are such that for any arc going from $\A$ to $\B$,
\begin{equation}\label{delta}
\langle \delta \A, v_\A\rangle=\langle\delta \B,v_\B\rangle,
\end{equation}
$v_\A$ and $v_\B$ being the initial and final velocity vectors on the arc. Theorems~1.1 and 1.4 are easily deduced from this observation, as we will see.

\medskip
{\bf Definition 1.5.} A {\it natural mechanical system} $({\cal M},g,U)$ is given by a manifold ${\cal M}$, called the {\it configuration space}, together with a
Riemannian bilinear form $g$ on ${\cal M}$  and  a {\it force function} $U:{\cal M}\to \R$  (see Arnold {\it et al.}\ [1]). The {\it kinetic energy} is $T$ with $2T=\langle g\dot q,\dot q \rangle$. Here the bracket $\langle g\dot q,\dot q \rangle$ is the duality pairing of vectors and covectors, $\dot q$ is the velocity vector, which is the time derivative of the  {\it configuration} $q\in {\cal M}$. The bilinear form $g$ is seen as a symmetric linear map which sends a vector $v$ tangent to ${\cal M}$ at a point $q$ to a covector $gv$ at $q$. The covector $g\dot q$ is called the {\it momentum}. Indeed, it is traditionally called the linear momentum in the case of a moving particle.

\medskip
{\bf Remark 1.6.} In most contexts one considers a unique Riemannian form $g$ on ${\cal M}$ and omits it in the notation. To omit it, one {\it identifies} the vector $v$ with the covector $gv$. The map $g$ is then the identity. The duality pairing becomes the inner product of vectors. Formulae (\ref{deltaS}) and (\ref{delta}) are written with this convention. In the context of \S \ref{ConC} there will be two Riemannian forms on ${\cal M}$, and we will not identify vectors and covectors. 

\medskip
In Theorems 1.1 and 1.4, the natural mechanical system is the Kepler problem, the manifold is the Euclidean plane with the origin $\O$ removed, the metric $g$ is the Euclidean bilinear form, and the force function is $U(q)=1/r$.

\medskip
{\bf Definition 1.7.} In a natural mechanical system, an {\it arc} is a part of any orbit which corresponds to a compact interval of time. The {\it ends} of an arc are the initial position $\A\in{\cal M}$ and the final position $\B\in{\cal M}$, which we do not assume to be distinct. The initial time $t_\A\in\R$ and the final time $t_\B\in\R$ should satisfy $t_\A<t_\B$.

\medskip
A natural mechanical system is a {\it Lagrangian system} with Lagrangian $L=T+U$. The Lagrangian action $S$ is stationary on any arc, compared with the paths with the same ends $\A$ and $\B$, same $t_\A$ and same $t_\B$. The energy $H=T-U$  is constant along any orbit.
A natural mechanical system is also a {\it Hamiltonian system}. The covector $p=g\dot q$ is the variable conjugated to the position $q$.
The {\it Hamiltonian} is $H(p,q)=\langle p,g^{-1} p\rangle/2-U(q)$ and the equations of motion are $\dot q=\partial H/\partial p$ and $\dot p=-\partial H/\partial q$. Guided by equation (\ref{delta}), we propose a new definition.

\medskip
{\bf Definition 1.8.} A {\it Lambert vector} at $(\A,\B)$ is a  vector $X$ tangent to ${\cal M}\times {\cal M}$ such that for any arc from $\A$ to $\B$,
\begin{equation}\label{Lv}
\langle g v_\A,\partial_X \A\rangle=\langle g v_\B,\partial_X \B\rangle
\end{equation}
where $v_\A$ and $v_\B$ are the initial and final velocities on the arc. A {\it Lambert vector field} is a vector field on ${\cal M}\times {\cal M}$ made of Lambert vectors.

\medskip
{\bf Notation.} Here we replace the traditional symbol $\delta$, used in (\ref{delta}) and widely used in this context since Lagrange and Hamilton, by $\partial_X$, which means differentiation in the direction of $X$. We will consider both symbols as synonyms. The vector $X$ tangent to ${\cal M}\times {\cal M}$ may be thought of as the pair $(\partial_X\A,\partial_X\B)$. We will denote by $\I$ an open interval of $\R$.

\medskip
{\bf Definition 1.9.} A {\it Lambert path} is a smooth path $\I\to {\cal M}\times {\cal M}$, $s\mapsto (\A_s,\B_s)$ such that for any $s$, the pair $(\d \A_s/\d s,\d \B_s/\d s)$ forms a Lambert vector.

\medskip
{\bf Paths in the space of arcs.} Smooth changes of arcs are smooth paths $\I\to {\cal A}$, $s\mapsto \Gamma_s$ in the space of arcs ${\cal A}$. Such a path induces a path  $\I\to {\cal M}\times {\cal M}$, $s\mapsto (\A_s,\B_s)$  in the space of pairs of ends.  Setting $\Delta t=t_\B-t_\A$, formula (\ref{deltaS}) now reads
\begin{equation}\label{ds}
\frac{\d S}{\d s}\Big |_{H=h}=\langle \frac {\d\B_s}{\d s},gv_{\B_s}\rangle-\langle \frac {\d\A_s}{\d s}, gv_{\A_s}\rangle-\frac{\d(\Delta t)_s}{\d s}h.
\end{equation}

\medskip
{\bf Proposition 1.10.} During a smooth change of arc, if the ends $(\A_s,\B_s)$ follow a Lambert path and if the elapsed time $(\Delta t)_s$ is constant, then the Lagrangian action $S$ is also constant.

\medskip
{\bf Proof.} As $(\d \A_s/\d s,\d \B_s/\d s)$ is a Lambert vector, the first two terms in (\ref{ds}) cancel each other out. As $(\Delta t)_s$ is constant, the last term also vanishes.  Thus $\d S/\d s=0$. \qed

\medskip
{\bf Remark 1.11.} Can the hypothesis of Proposition 1.10 be fulfilled? Given a Lambert path $s\mapsto (\A_s,\B_s)$, given an arc at $s=0\in{\cal I}$ with ends $(\A_0,\B_0)$, can we find a path of arcs $s\mapsto \Gamma_s$ respecting both conditions, that  the ends of $\Gamma_s$ are $\A_s$ and $\B_s$, and that the elapsed time $(\Delta t)_s$ is constant? The answer is yes, locally, if the mechanical system and the arc satisfy this {\it non-degeneracy condition}:  the flow of the mechanical system during the time $(\Delta t)_0$ has a non-zero Jacobian determinant at $\A_0$. Such condition is satisfied in most situations, but is not satisfied in the important examples of a Keplerian arc corresponding to a full period, i.e.\ with $\A=\B$, and of an arc of the Hooke problem such that the midpoint of $\A$ and $\B$ is $\O$.

\medskip
{\bf Proposition 1.12.} During a smooth change of arc, if the ends $(\A_s,\B_s)$ follow a Lambert path, if the non-degeneracy condition of remark~1.11 is satisfied at each $\A_s$ and if the elapsed time $(\Delta t)_s$ is constant, then the energy $H$ is also constant.

\medskip
{\bf Proof.} An arc satisfying the non-degeneracy condition belongs to a family of arcs with the same ends, parametrized by $\Delta t$. Formula (\ref{ds}) applied to this family gives $\d S/\d(\Delta t)=-H$. When $(\A,\B)$ changes this family changes. If $(\A,\B)$ follows a Lambert path, $S$ remains the same function of $\Delta t$ according to Proposition~1.10. Consequently $H=-\d S/ \d(\Delta t)$ also remains the same function of $\Delta t$, i.e.\ is constant when $(\A,\B)$ is changing. \qed

\medskip
Two variants of the previous two propositions are obtained from this variant of (\ref{deltaS}), also given by Hamilton:
 \begin{equation}
\delta w=\langle \delta \B,v_\B\rangle-\langle \delta \A, v_\A\rangle+\Delta t \cdot \delta H.
\end{equation}

\medskip
{\bf Proposition 1.13.} During a smooth change of arc, if the ends $ (\A_s,\B_s)$ follow a Lambert path and if the energy $H$ is constant, then the Maupertuis action
$w$ is also constant. 

\medskip
{\bf Proposition 1.14.} During a smooth change of arc, if the ends $(\A_s,\B_s)$ follow a Lambert path, if the non-degeneracy condition of remark~1.11 is satisfied at each $\A_s$ and if the energy $H$ is constant, then the elapsed time $\Delta t$ is also constant.

\medskip
The proofs are similar. The non-degeneracy condition does not imply that $H$ is a local parameter of the family of arcs with ends $\A$ and $\B$, but we can still argue with $\Delta t$ as the parameter of this family. We do not insist on these proofs in the general framework, based on the non-degeneracy condition, since they only apply to rather simple systems, for which the statements may easily be improved. For example, Theorem~1.1 is Proposition~1.14 in the case of the Kepler problem, but it also applies to the degenerate case $\A=\B$.

In the same example of the Kepler problem, here in the plane $\O xy$, we set $\A=(x_\A,y_\A)$ and $\B=(x_\B,y_\B)$. Then $(\delta\A,\delta\B)=(-y_\A,x_\A,-y_\B,x_\B)$ forms a Lambert vector. To check this, it is enough to note that (\ref{delta}) expresses the conservation of the angular momentum. The four propositions are trivial consequences of the rotational invariance of the problem. For a natural mechanical system, the situation is the same as in the particular case of the Kepler problem, according to the following general proposition.

\medskip
{\bf Proposition 1.15.} Let $({\cal M}, g, U)$ be a natural mechanical system, ${\cal L}$ denote the Lie derivative and $v=\dot q$ be the velocity vector. A vector field $Z$ on ${\cal M}$ satisfies $\langle \d U,Z\rangle=0$ and ${\cal L}_Zg=0$ if and only if $\langle gZ,v\rangle$ is a first integral of the system.

\medskip
{\bf Proof.} Let $\rho=gZ$, let $\nabla$ be the Levi-Civita connection of $g$. Recall that for any tangent vectors $X$, $\langle \nabla_X \rho,X\rangle=\langle {\cal L}_Zg,X\otimes X\rangle$. Then
\begin{equation}
\partial_v\langle gZ,v\rangle=\langle \nabla_v\rho,v\rangle+\langle\rho,\nabla_v v\rangle=\langle {\cal L}_Zg,v\otimes v\rangle+\langle Z,\d U\rangle.
\end{equation}
If $\langle gZ,v\rangle$ is a first integral, this is zero for all velocity vectors $v$. Then ${\cal L}_Zg$ is zero and $\langle Z,\d U\rangle=0$. The only if part also follows from this formula. \qed

\medskip
{\bf Definition 1.16.} A {\it natural symmetry} of a natural mechanical system is a group of isometries of the configuration space $({\cal M},g)$ which preserves the force function $U$. A {\it trivial Lambert vector} at $(\A,\B)$ is a pair of vectors $(Z|_\A,Z|_\B)$, where $Z$ is a vector field on ${\cal M}$ generating a continuous natural symmetry. We say that a natural mechanical system possesses the {\it Lambert-Hamilton property} if it admits a non-trivial Lambert vector field.
\medskip

The vector field $Z$ in Definition 1.16 satisfies the conditions in Proposition 1.15. In particular, $Z$ is a {\it Killing vector field} of the Riemannian manifold $({\cal M},g)$. See e.g.\ O'Neill [1]. In  the rest of this paper, we will concentrate on the following

\medskip
{\bf General question.} Does a given natural mechanical system have the Lambert-Hamilton property?

\medskip
We will successively present six systems which positively answer the general question: the Kepler problem and the Hooke problem, and their analogues on a space of constant positive curvature and on a space of constant negative curvature. The property also extends to the same systems with a repulsive force instead of attractive. Further possibilities are obtained by considering pseudo-Riemannian forms $g$.

\section{Flat configuration space}

\subsection{Classical Lambert's theorem}

To rediscover the statement of Lambert's theorem, we consider the Kepler problem in the plane $\O xy$, defined by the system of ordinary differential equations

\begin{equation}
\ddot x=-{x\over r^3}, \qquad \ddot y=-{y\over r^3},\qquad \hbox{where } r=\sqrt{x^2+y^2},
\end{equation}
and address the above general question. We look for a Lambert vector at each $(\A,\B)$. Any Lambert path $s\mapsto (\A_s,\B_s)$ may be composed at each $s$ with a rotation around $\O$, thus producing a new Lambert path. We choose a normalization which eliminates the rotational symmetry of the system: we decide that $y_\A=y_\B$ for all $(\A,\B)$ in the Lambert path.\footnote{Indeed the reasoning and the computation remain valid under the weaker hypothesis that $y_\A-y_\B$ remains constant along the Lambert path, and that $x_\A\neq x_\B$. This is why we keep the two symbols $y_\A$ and $y_\B$ during the computation, even if we assume $y_\A=y_\B$.}

To get the Lambert vectors, we need to determine the velocities $(v_\A,v_\B)$ at the ends $(\A, \B)$ of an arc of orbit. Three classical conserved quantities, the angular momentum $C=x\dot y-y\dot x$, and both coordinates of the eccentricity vector, $\alpha=x/r-\dot yC$, $\beta=y/r+\dot xC$, give expressions for these velocities. They give:
\begin{equation}
Cv_\A=\bigl(\beta-{y_\A\over r_\A},-\alpha+{x_\A\over r_\A}\bigr),\qquad Cv_\B=\bigl(\beta-{y_\B\over r_\B},-\alpha+{x_\B\over r_\B}\bigr).
\end{equation}
 Here we use the notation $r_\A=\sqrt{x_\A^2+y_\A^2}$, $r_\B=\sqrt{x_\B^2+y_\B^2}$. If $C\neq 0$,  condition (\ref{Lv}) for a Lambert vector $X$ reads
$$0=-\bigl({y_\B\over r_\B}-\beta\bigr)\partial_X x_\B+\bigl({x_\B\over r_\B}-\alpha\bigr)\partial_X y_\B+\bigl({y_\A\over r_\A}-\beta\bigr)\partial_X x_\A-\bigl({x_\A\over r_\A}-\alpha\bigr)\partial_X y_\A.$$
Together with the normalization $\partial_X y_\A=\partial_X y_\B$ this gives
\begin{equation}\label{L}
0=-\bigl({y_\B\over r_\B}-\beta\bigr)\partial_X x_\B+\bigl({x_\B\over r_\B}-{x_\A\over r_\A}\bigr)\partial_X y_\A+\bigl({y_\A\over r_\A}-\beta\bigr)\partial_X x_\A.
\end{equation}
This should be true for any orbit passing through $\A$ and $\B$. The easy identity $\alpha x+\beta y=r-C^2$ is the equation of the orbit, which is a conic section with focus at the origin $\O$ and directrix with equation $\alpha x+\beta y+C^2=0$. Of the three parameters in the equation of the orbit, $\alpha$, $\beta$, $C$, only $\beta$ appears in (\ref{L}). Therefore, this equation is true for any orbit if it is true for any $\beta$, and it is true for any $\beta$ if:
\begin{equation}\label{L1}
0=-{y_\B\over r_\B}\partial_X x_\B+\bigl({x_\B\over r_\B}-{x_\A\over r_\A}\bigr)\partial_X y_\A+{y_\A\over r_\A}\partial_X x_\A\quad\hbox{and}\quad 0=\partial_X x_\B-\partial_X x_\A.
\end{equation}
The second equation (\ref{L1}) implies that the segment $\A\B$ is translated when following the Lambert path. The first equation (\ref{L1}) becomes:
\begin{equation}0=-\bigl({y_\B\over r_\B}-{y_\A\over r_\A}\bigr)\partial_X x_\A+\bigl({x_\B\over r_\B}-{x_\A\over r_\A}\bigr)\partial_X y_\A.\end{equation}
Being a linear form in the two unknowns $(\partial_Xx_\A,\partial_Xy_\A)$, this equation possesses solutions. Consequently, {\it there exists a non-trivial Lambert vector field}. We choose an $X$ by choosing a normalization:
\begin{equation}\label{Th}
\partial_X x_\A=\partial_X x_\B={x_\B\over r_\B}-{x_\A\over r_\A},\qquad \partial_X y_\A=\partial_X y_\B={y_\B\over r_\B}-{y_\A\over r_\A}.
\end{equation}
This expression defines a $(\partial_Xx_\A,\partial_Xy_\A)$ which is equivariant by rotation of $(\A,\B)$, and consequently valid even if $y_\A\neq y_\B$.

\medskip
{\bf Proposition 2.1.} The  vector field $X$ defined on ${\cal M}\times{\cal M}$ by system (\ref{Th}), where ${\cal M}=\R^2\setminus\{\O\}$, is a non-trivial Lambert vector field for the Kepler problem in the plane $\R^2=\O xy$.

\medskip
{\bf The invariants.} There remains to find an invariant for the vector field $X$ defined by (\ref{Th}). We compute
$$\partial_X r_\A={1\over r_\A}\Bigl(\bigl({x_\B\over r_\B}-{x_\A\over r_\A}\bigr)x_\A+\bigl({y_\B\over r_\B}-{y_\A\over r_\A}\bigr)y_\A\Bigr)={x_\A x_\B+y_\A y_\B\over r_\A r_\B}-1.$$
$$\partial_X r_\B={1\over r_\B}\Bigl(\bigl({x_\B\over r_\B}-{x_\A\over r_\A}\bigr)x_\B+\bigl({y_\B\over r_\B}-{y_\A\over r_\A}\bigr)y_\B\Bigr)=1-{x_\A x_\B+y_\A y_\B\over r_\A r_\B}.$$
Thus, $r_\A+r_\B$ is invariant along the Lambert paths generated by the Lambert vector field $X$. 
So, $X$ is characterized by two conditions: the pair $(\A,\B)$ is translated and $r_\A+r_\B$ remains constant. The general Lambert paths for the Kepler problem are compositions of arbitrary rotations with the paths generated by $X$. They are characterized by the invariance of $d(\A,\B)$ and  $d(\O,\A)+d(\O,\B)$.

Proposition 1.14 together with this description of the Lambert paths is essentially Theorem~1.1.  We thus obtain, for the first time, Lambert's theorem as the answer to a general question, which appears to be easily solvable.

Proposition 1.14 has a non-degeneracy hypothesis which we can avoid as follows. To guarantee that a given arc is not isolated among the arcs from $\A$ to $\B$, we observe that the family of orbits passing through $\A$ and $\B$ corresponds to parameters $(\alpha, \beta, C)$ such that $\alpha x_\A+\beta y_\A+C^2=r_\A$ and $\alpha x_\B+\beta y_\B+C^2=r_\B$. The three parameters $(\alpha, \beta, C^2)$ are constrained by two affine conditions only. This proves that the arc is not isolated. The same argument shows that the particular case $\A=\B$ is not an exception, although the non-degeneracy condition is not satisfied. The case where $\A$ and $\B$ are distinct, but on a same ray, is not an exception to Theorem~1.1 either. It should be considered as a limit in the above reasonings, where we assumed $C\neq 0$. See Albouy [3] for a discussion of rectilinear orbits and Lambert's theorem, and the extension of the Kepler problem after collision.
The case $\O=\A$ is not an exception either.

\subsection{The Hooke problem in the plane}  The planar harmonic oscillator, which we call the Hooke problem, is defined in the plane $\O xy$ by the system of ordinary differential equations $\ddot x=-x$, $\ddot y=-y$. We address the general question. As expected, the study is  simpler than the above study of the Kepler problem.

Let us normalize, using the rotations, with the condition $y_\A=y_\B$. The variables $x$ and $y$ are separated in the differential system, each having a periodic motion of period $2\pi$. Considering the variable $y$ alone, we observe that for any orbit passing through $\A$ and $\B$, we have $\dot y_\A=-\dot y_\B$ or $\dot y_\A=\dot y_\B$.  In the first case, $\partial_X\A=(0,1)$, $\partial_X\B=(0,-1)$ define a Lambert vector $X$, since $\langle \partial_X \A,v_\A\rangle=\langle \partial_X \B,v_\B\rangle$ for any such orbit. To make a Lambert vector field defined everywhere, not only for $y_\A=y_\B$, we  propose the following formulae, which are equivariant by rotation of $(\A,\B)$:
\begin{equation}\label{H}
\partial_X \A=(y_\A-y_\B,x_\B-x_\A),\qquad\partial_X \B=(y_\B-y_\A,x_\A-x_\B).
\end{equation}
In the second case, since $(y_\A,\dot y_\A)=(y_\B,\dot y_\B)$, the interval of time must be a multiple of the period $2\pi$. Consequently $\A=\B$. Formula (\ref{H}) is still correct in this case, since it gives zero, which is a Lambert vector.

\medskip
{\bf Proposition 2.2.} The  vector field $X$ defined on ${\cal M}\times{\cal M}$ by system (\ref{H}), where ${\cal M}=\R^2=\O xy$, is a non-trivial Lambert vector field for the Hooke problem.
\medskip

{\bf The invariants.} They may be presented in several ways. The following computation will be useful in \S \ref{HooS}: $r_\A\partial_X r_\A=-x_\A y_\B+x_\B y_\A$ and $r_\B\partial_X r_\B=-x_\B y_\A+x_\A y_\B$, thus $r_\A\partial_X r_\A+r_\B\partial_X r_\B=0$. This shows that $r_\A^2+r_\B^2$ is invariant along a Lambert path of the Hooke problem. Now consider the chord $\|\A\B\|$.
\begin{equation}\partial_X \|\A\B\|^2=4(x_\A-x_\B)(y_\A-y_\B)+4(y_\A-y_\B)(x_\B-x_\A)=0.\end{equation}
Now denote by $\A'$ the opposite of point $\A$, with coordinates $(-x_\A,-y_\A)$. We call $\|\A'\B\|$ the {\it antichord}. We also have
\begin{equation}\partial_X \|\A'\B\|^2=0.\end{equation}
Thus, we have proved:

\medskip
{\bf Theorem 2.3.} Consider the Hooke problem, i.e.\ the harmonic oscillator in a Euclidean plane, with fixed centre at $\O$. The time required to reach a point $\B$ 
from a point $\A$ with a given total energy $H$ does not vary if we continuously change $\A$ and $\B$ in such a way that the distances $d(\A,\B)$ and $d(\A',\B)$, where $\A'$ is such that $\O$ is the midpoint of $(\A,\A')$, remain constant.

\medskip
{\bf Remark 2.4.} If two arcs of the Hooke problem have the same elapsed time, they are mapped one onto the other by some linear map. Consequently, an arc changed by following a Lambert path with a constant energy is a linear transformation of the initial arc. A similar property exists in the Kepler problem. When following a Lambert path with constant energy $H$, a Keplerian arc remains in the same affine class (see Albouy [3]).

\section{Non-zero constant curvature and Appell's projection}
\label{ConC}

\subsection{Appell's central projection and Lambert's theorem}
In the above argument for the Kepler problem in the plane, we deduce the existence of a Lambert vector before computing its expression. We express the definition of a Lambert vector as a linear system (\ref{L1}) with two equations in four variables. This system is underdetermined in such a way that the rotation cannot be the only solution and there should exist another Lambert vector. This argument gives the following theorem as a development. The theorem gives in turn further examples of systems with a Lambert-Hamilton property.

\medskip
{\bf Theorem 3.1.} If two {\it systems with radial reaction} correspond with each other by Appell's projection and if both are natural systems, then there is an invertible linear map from the Lambert vectors at $(\A,\B)$ to the Lambert vectors at the image of $(\A,\B)$. This map sends the trivial Lambert vectors onto the trivial Lambert vectors.

\medskip
{\bf Remark 3.2.} Except in trivial cases, this linear map is not the push-forward by Appell's central projection and consequently the Lambert paths do not coincide through this projection.

\medskip
{\bf Description of Appell's projection.} The {\it systems with radial reaction} are the systems which are projected one onto the other by Appell's projection. In the case of a zero reaction, they reduce to systems defined by a force field on a flat space. Let us begin by explaining Appell's projection in this case. In a finite dimensional real vector space ${\cal W}$ we consider two hyperplanes which do not pass through the origin $\Omega$ of ${\cal W}$, a connected open set ${\cal U}$ in the first hyperplane, and a connected open set ${\cal V}$ in the second. We assume that ${\cal U}$ and ${\cal V}$ correspond with each other by central projection with centre $\Omega$. More precisely, the correspondence should be through half lines and not lines, i.e.\ each ray emanating from $\Omega$ and passing through a point of ${\cal U}$ also passes through a point of  ${\cal V}$, and the same is true if we exchange ${\cal U}$ and ${\cal V}$. Suppose that a dynamics on ${\cal U}$ is defined by a vector field $\phi$, through the Newton equation $\ddot q=\phi|_q$. We call $\phi$ a {\it force field}. Then a corresponding dynamics on ${\cal V}$ is defined by a force field $\psi$ on ${\cal V}$. The rules are (see Albouy [2]):

{\bf First rule.} The moving point $q_{\cal U}\in {\cal U}$ is sent onto the moving point $q_{\cal V}\in {\cal V}$ by central projection: there is a factor $\gamma> 0$ depending only on $q_{\cal U}$  such that $q_{\cal V}=\gamma q_{\cal U}$.

{\bf Second rule.} The velocity vector $v_{\cal V}$ of $q_{\cal V}$ is the push-forward of the velocity vector $v_{\cal U}$ of $q_{\cal U}$ by the central projection, divided by $\gamma^2$.

{\bf Third rule.} The force vector $\psi$ at $q_{\cal V}$ is the push-forward of the  force vector $\phi$ at $q_{\cal U}$ by the central projection, divided by $\gamma^4$.

The open sets ${\cal U}$ and ${\cal V}$ are called {\it screens}, since a moving ray of light emanating from $\Omega$ draws the orbits of the moving points $q_{\cal U}$ and $q_{\cal V}$ on each screen. The time parameters on each screen do not coincide: the division by $\gamma^2$ in the second rule defines a change of time.

The screens ${\cal U}$ and ${\cal V}$ are flat, but Appell's projection also admits curved screens, which are hypersurfaces of the vector space ${\cal W}$. Such a screen should always intersect the rays from $\Omega$ transversally. When the screen is not flat, the law of dynamics is specified by adding to the force field $\phi$ a vectorial {\it reaction} $\lambda q$:
\begin{equation}
\ddot q=\phi|_q+\lambda q,\end{equation} In the context of Appell's projection, this reaction is always directed radially: the radial vector $q$ is multiplied by a scalar $\lambda$ which is uniquely determined in such a way that $q$ remains on the curved screen all along the motion. This radial reaction was introduced in Albouy [1].

\medskip
{\bf Proof of Theorem 3.1}. We consider a given natural system $({\cal M},g,U)$, where the configuration space ${\cal M}$ is a screen, and project it by Appell's projection on another configuration space which is another screen. We assume that the resulting system is a natural mechanical system: the force field defined by the third rule is for some Riemannian metric the gradient of some force function. We pull back this system by the projection, for an easier comparison of both systems. We call the new natural system $({\cal M},\hat g,\hat U)$. Here we  assume that both systems have the same domain ${\cal M}$, which we can always obtain by restricting to a common domain. Consider two points $\A$ and $\B$ in ${\cal M}$. Let $\gamma_\A>0$ and $\gamma_\B>0$ the factors defined by the first rule. For any arc from $\A$ to $\B$ starting with velocity $v_\A$ and arriving with velocity $v_\B$, there is a motion of the new system on the same arc, with initial and final velocities $(\hat v_\A,\hat v_\B)=(\gamma_\A^{-2} v_\A,\gamma_\B^{-2} v_\B)$, according to the second rule. Now, in order to get the condition (\ref{Lv}) for a Lambert vector $Y=(\partial_Y \A,\partial_Y \B)$, namely, $\langle \partial_Y \A, \hat g \hat v_\A\rangle=\langle \partial_Y \B, \hat g \hat v_\B\rangle$ from the condition $\langle \partial_X \A, g v_\A\rangle=\langle \partial_X \B, g v_\B\rangle$, it is enough to define the new Lambert vector $Y$  through the relations $\partial_Y \A=\gamma_\A^2\, \hat g^{-1} g\, \partial_X \A$ and $\partial_Y \B=\gamma_\B^2\, \hat g^{-1} g \,\partial_X \B$. This is the invertible linear map of the Theorem. To obtain the result about the trivial Lambert vectors, consider a generator $Z$ of a natural symmetry group of $(g,U)$. By Proposition~1.15, $\langle Z,gv\rangle$ is a first integral of System $(g,U)$. Consequently, $\langle Z,gv\rangle=\langle \hat Z,\hat g\hat v\rangle$, where $\hat Z=\gamma^2\, \hat g^{-1} g \, Z$, is a first integral of System $(\hat g,\hat U)$. By Proposition~1.15, $\hat Z$ is a natural symmetry of $(\hat g,\hat U)$. Trivial Lambert vectors $(Z|_\A,Z|_\B)$ are mapped onto trivial Lambert vectors $(\hat Z|_\A,\hat Z|_\B)$. \qed

\subsection{Application of Appell's central projection}
The Kepler problems on constant curvature spaces, introduced by Paul Serret [1] in 1859 and Wilhelm Killing [1] in 1885, are natural systems which, according to Appell [2], are obtained from the usual Kepler problem by central projection.   The same is true for the Hooke problems on constant curvature spaces (see Albouy [2]). We will recall the precise constructions in the next sections. Theorem~3.1 gives the following proposition.

\medskip
{\bf Proposition 3.3.} The Kepler problem and the Hooke problem on constant curvature surfaces admit non-trivial Lambert vector fields.

\medskip
{\bf Remark 3.4.} In Theorem 3.1 the domains of both systems correspond with each other. But in the example of the spherical Kepler problem, Appell's projection maps the usual Kepler problem onto just a hemisphere, the ``upper'' hemisphere, centred at the fixed centre $\O$. Proposition~3.3 would apply only to the restriction of the spherical Kepler problem to the upper hemisphere, while we claim it applies to the full spherical Kepler problem. There are indeed two distinct objections, which are both easy to resolve.

The first is that the Lambert vectors are defined through the velocities $(v_\A,v_\B)$ of all the arcs going from $\A$ to $\B$. Increasing the domain does create new arcs, which in turn create new velocities. The definitions of Lambert vectors for both domains do not coincide. In the spherical Kepler problem, if $\A$ and $\B$ are in the upper hemisphere, some arcs from $\A$ to $\B$ do cross the equator. But such an arc of spherical ellipse has, up to a change of signs, the same velocities $(v_\A,v_\B)$ as the complementary arc, which is contained in the hemisphere. So the Lambert vectors do correspond with each other in both systems, spherical and flat.

The second objection is that Theorem 3.1 defines the Lambert vectors fields only on the upper hemisphere. But these fields are analytic, and the formula for them produces an analytic extension to the lower hemisphere, satisfying the required equations.

The spherical Hooke problem is singular on the equator, and defined only on the upper hemisphere. Appell's projection maps it completely onto the standard Hooke problem.

\subsection{Kepler on the sphere}

We consider a conservative central force problem in the plane $\O xy$, where the force is directed toward a centre $\O$. The force function $U$ is a function of the distance $r$. We embed the plane in a three-dimensional Euclidean space and consider the unit sphere tangent to the plane at $\O$. By calling $\Omega$ its centre, and by projecting centrally toward $\Omega$, according to Appell's rule, the conservative central force problem is transformed into a natural mechanical system on the sphere. The new force function $\hat U$ is obtained by projecting $U$. In the Kepler problem, $U=1/r$, and, if $\theta$ is the angle from the pole $\O$,
\begin{equation}\label{NS}
\hat U={1\over \tan\theta}.
\end{equation}
The kinetic energy is changed. Let the old kinetic energy $T$ be such that $2T=\dot x^2+\dot y^2$. The new kinetic energy, expressed on the same plane with the same variables $x$ and $y$, is
\begin{equation}\label{TS}
\hat T={\dot x^2+\dot y^2+(x\dot y-y\dot x)^2\over 2(1+x^2+y^2)^2}.
\end{equation}
The $\gamma$ factor of the first rule is $\gamma=(1+x^2+y^2)^{-1/2}$. The matrix $g$ in the proof of Theorem~3.1 is the identity, while
\begin{equation}\hat g=\gamma^{4}\pmatrix{1+y^2&-xy\cr -xy&1+x^2}.\end{equation}
Then 
\begin{equation}\label{Dd}
\hat g^{-1}=\gamma^{-2}\pmatrix{1+x^2&xy\cr xy&1+y^2}.
\end{equation}
We set
\begin{equation}G_\A=\pmatrix{1+x_\A^2&x_\A y_\A\cr x_\A y_\A&1+y_\A^2},\quad G_\B=\pmatrix{1+x_\B^2&x_\B y_\B \cr x_\B y_\B&1+y_\B^2}.\end{equation}
Since the factors $\gamma_\A^2$ and $\gamma_\B^2$ in the proof of Theorem~3.1 simplify the $\gamma^{-2}$ factor in (\ref{Dd}), the invertible linear map in this Theorem appears in matrix form as
\begin{equation}\label{M}
 \partial_Y\A=(\partial_X\A)G_\A,\qquad \partial_Y\B=(\partial_X\B)G_\B,
 \end{equation}
 where, for the usual Kepler problem, $X$ is defined by formula (\ref{Th}), which is
$$\partial_X \A=\partial_X \B=\bigl({x_\B\over r_\B}-{x_\A\over r_\A}, {y_\B\over r_\B}-{y_\A\over r_\A}\bigr).$$
The matrix
product (\ref{M}) gives the Lambert vector $Y$ from the previous Lambert vector $X$ of the planar case. We do not need to expand this product. We compute the invariants of $Y$. We have $$r_\A \partial_Y r_\A=x_\A \partial_Y x_\A+y_\A \partial_Y y_\A=(\partial_X\A)G_\A\pmatrix{x_\A\cr y_\A}$$
$$=(1+r_\A^2)(\partial_X\A)\pmatrix{x_\A\cr y_\A}=(1+r_\A^2)r_\A \partial_X r_\A.$$
So \begin{equation}\partial_Y r_\A=(1+r_\A^2) \partial_X r_\A,\quad \partial_Y r_\B=(1+r_\B^2) \partial_X r_\B.\end{equation}
We have seen in the planar case that $\partial_Xr_\A+\partial_Xr_\B=0$. So
\begin{equation}{\partial_Y r_\A\over 1+r_\A^2}+{\partial_Y r_\B\over 1+r_\B^2}=0.\end{equation}
If we call $\hat\A$ and $\hat\B$ the central projections of $\A$ and $\B$ on the sphere, $\theta_\A$ the angle $\O\hat\A$  and $\theta_\B$ the angle $\O\hat\B$, then $r_\A=\tan\theta_\A$ and
$r_\B=\tan\theta_\B$. The above relation becomes $\partial_Y\theta_\A+\partial_Y\theta_\B=0$. {\it Thus, $\theta_\A+\theta_\B$ is an invariant for $Y$.} This suggests to ask if the angle $\hat\A\hat\B$ on the sphere, denoted by $\theta_{\A\B}$, is invariant. Since $\cos\theta_{\A\B}=(1+r_\A^2)^{-1/2}(1+r_\B^2)^{-1/2}(1+x_\A x_\B+y_\A y_\B)$, we compute $\partial_Y(1+r_\A^2)=2(1+r_\A^2)r_\A\partial_X r_\A$, $\partial_Y(1+r_\B^2)=2(1+r_\B^2)r_\B\partial_X r_\B$ and $$\partial_Y(1+x_\A x_\B+y_\A y_\B)=(\partial_X\A)G_\A\pmatrix {x_\B\cr y_\B}+(\partial_X\B)G_\B\pmatrix {x_\A\cr y_\A}$$
\begin{equation}\label{s}=(\partial_X\A)\pmatrix{(1+x_\A^2)x_\B+x_\A y_\A y_\B+(1+x_\B^2)x_\A+x_\B y_\B y_\A\cr x_\A y_\A x_\B+(1+y_\A^2)y_\B+ x_\B y_\B x_\A+(1+y_\B^2)y_\A}
\end{equation}
$$=(\partial_X\A)\pmatrix{(x_\A+x_\B)(1+x_\A x_\B+y_\A y_\B)\cr (y_\A+y_\B)(1+x_\A x_\B+y_\A y_\B)}$$
$$=(1+x_\A x_\B+y_\A y_\B)(r_\A \partial_X r_\A+r_\B\partial_X r_\B).$$
Then  $\partial_Y(\log \cos\theta_{\A\B})=-r_\A\partial_X r_\A-r_\B\partial_X r_\B+(r_\A \partial_X r_\A+r_\B\partial_X r_\B)=0$.
{\it Thus, $\theta_{\A \B}$ is also an invariant for $Y$}.

What we have proved using the central projection may now be stated without mention of this projection. We will use the notation $\A$, $\B$ and $H$ for what was denoted in the above proof $\hat\A$, $\hat\B$ and $\hat H=\hat T -\hat U$.

\medskip
{\bf Theorem 3.5.} Consider the Kepler problem on a unit sphere: a particle moves on the sphere under a force field which is the gradient of the force function $1/\tan\theta$, where $\theta$ is the angle (or geodesic distance) from a fixed centre $\O$. The time required to reach a point $\B$ from a point $\A$ with a given total energy $H$ does not vary if we continuously change $\A$ and $\B$ in such a way that the geodesic distances $d(\A,\B)$ and  $d(\O,\A)+d(\O,\B)$ remain constant.

\subsection{Kepler in negative constant curvature}

The model for the hyperbolic plane is a sheet of the two-sheeted unit hyperboloid in Minkowski space, which is sometimes called the pseudosphere. Appell's projection maps it onto a disk. The computations are the same with the usual changes, such as $r_\A=\tanh \theta_A$ and \begin{equation}\hat T={\dot x^2+\dot y^2-(x\dot y-y\dot x)^2\over 2(1-x^2-y^2)^2}.\end{equation}

{\bf Theorem 3.6.} Theorem 3.5 is still true if we replace the unit sphere by a unit pseudosphere, i.e.\ by the hyperbolic space, and the force function $1/\tan\theta$ by $1/\tanh s$, where $s$ is the geodesic distance from a fixed centre $\O$.

\subsection{The Hooke problem on the sphere}
\label{HooS}

This problem is the central projection on the unit sphere of the usual Hooke problem. The force function of the latter problem, $U=-r^2/2$, is conserved by the projection, giving
\begin{equation}
\hat U=-{1\over 2}\tan^2\theta
\end{equation}
on the sphere, $\theta$ being the angle from the fixed centre $\O$. The ``equator'' corresponding to the ``pole''  $\O$ is a singularity for $\hat U$. We restrict our study to a hemisphere. The kinetic energy is (\ref{TS}). We get the Lambert vector by applying the relation (\ref{M}) to the expressions (\ref{H}). The invariance relation $r_\A\partial_X r_\A+r_\B\partial_X r_\B=0$ is modified as in the Keplerian case, giving 
\begin{equation}{r_\A\partial_Yr_\A\over 1+r_\A^2} +{r_\B\partial_Y r_\B\over 1+r_\B^2}=0,\end{equation} 
which in turn gives the invariant $(1+r_\A^2)(1+r_\B^2)$. Another invariant, the inverse of the square root of this one, is $\cos\theta_\A\cos\theta_\B$. As $\partial_X \A=-\partial_X \B$, equation (\ref{s}) now has a minus sign:
$$\partial_Y(x_\A x_\B+y_\A y_\B)=(\partial_X\A)\pmatrix{(1+x_\A^2)x_\B+x_\A y_\A y_\B-(1+x_\B^2)x_\A-x_\B y_\B y_\A\cr x_\A y_\A x_\B+(1+y_\A^2)y_\B-x_\B y_\B x_\A-(1+y_\B^2)y_\A
}$$
$$=(y_\A-y_\B,x_\B-x_\A)\pmatrix{(x_\A-x_\B)(-1+x_\A x_\B+y_\A y_\B)\cr (y_\A-y_\B)(-1+x_\A x_\B+ y_\A y_\B)}=0.$$
Then $x_\A x_\B+y_\A y_\B$ is an invariant. Using the formula
\begin{equation}\cos\theta_{\A\B}={1+x_\A x_\B+y_\A y_\B\over \sqrt{1+r_\A^2}\sqrt{1+r_\B^2}}\end{equation}
as in the Keplerian case, and the above invariance of the denominator, we find that  {\it again, $\theta_{\A \B}$ is an invariant.}
We remark that if $\A'$ is obtained from $\A$ by a rotation by $\pi$ around the vertical axis, then
\begin{equation}\cos\theta_{\A'\B}={1-x_\A x_\B-y_\A y_\B\over \sqrt{1+r_\A^2}\sqrt{1+r_\B^2}}\end{equation}
is also invariant. In other words, the {\it antichord $\theta_{\A'\B}$ is invariant}. Clearly, the chord and the antichord form a basis of the invariants.

\medskip
{\bf Theorem 3.7.} Consider the Hooke problem on a hemisphere centred at the fixed centre $\O$. The time required to reach a point $\B$ 
from a point $\A$ with a given total energy $H$ does not vary if we continuously change $\A$ and $\B$ in such a way that the geodesic distances $d(\A,\B)$ and  $d(\A',\B)$, where $\A'$  is such that $\O$ is the midpoint of $\A\A'$, remain constant.

\subsection{The Hooke problem on the pseudosphere}

It is enough to apply the same changes as in the Keplerian case: $1+r_\A^2$ becomes $1-r_\A^2$, which is $\cosh^{-2}\theta_\A$.

\medskip
{\bf Theorem 3.8.} Theorem~3.7 is still true if we replace the unit hemisphere by a unit pseudosphere, i.e.\ by the hyperbolic space, and the force function $-(\tan^2\theta)/2$ by $-(\tanh^2 s)/2$, where $s$ is the geodesic distance from a fixed centre $\O$.

\bigskip
{\bf Acknowledgements.} We wish to thank Alain Chenciner, Christian Velpry and Camille Laurent-Gengoux for many stimulating comments. Many thanks to Niccol\`o Guicciardini for showing us the cited paragraphs of the {\it Treatise of Fluxions}. Thanks to the anonymous reviewers for their considerable help. Lei Zhao is supported by DFG ZH 605/1-1.

\section*{References}

 Albouy  A [1] Projective dynamics and classical gravitation, {\it Regular and Chaotic Dynamics}, 13 (2008), pp.\ 525--542

Albouy A [2] There is a projective dynamics, {\it EMS Newsletter}, 89 (2013), pp.\ 37--43

Albouy A [3]  Lambert's theorem: geometry or dynamics? {\it Celestial Mechanics and Dynamical Astronomy}, 131 (2019), 40

Appell P [1] De l'homographie en m{\'e}canique, {\it American Journal of Mathematics}, 12 (1890), pp.\ 103--114

Appell P [2] Sur les lois de forces centrales faisant d{\'e}crire {\`a} leur point d'appli\-cation une conique quelles que soient les conditions initiales, {\it American Journal of Mathematics}, 13 (1891), pp.\ 153--158

Arnold VI, Kozlov VV, Neishtadt, AI [1]  {\it Mathematical aspects of classical and celestial mechanics}, third edition, Springer 2007, p.\ 47

 Borisov AV, Mamaev IS [1] Rigid body dynamics in non-Euclidean spaces,
{\it Russian Journal of Mathematical Physics}, 23 (2016), pp.\ 431--454

 Borisov AV, Mamaev IS, Bizyaev IA [1] The Spatial Problem of 2 Bodies on a Sphere. Reduction and Stochasticity. {\it Regular and Chaotic Dynamics}, 21 (2016), pp.\ 556--580

Clifford WK [1] On the Free Motion under no Forces of a Rigid System in an $n$-fold Homaloid, {\it Proc.\ Lond.\ Math.\ Soc.\ }s1-7 (1876), pp.\ 67--70

 Darboux G [1] Remarque sur la Communication pr{\'e}c{\'e}dente, {\it Comptes Rendus  Acad.\  Sci.\ Paris}, 108 (1889), pp.\ 449--450

 Goursat E [1] Les transformations isogonales en M{\'e}canique, {\it Comptes Rendus  Acad.\  Sci.\ Paris}, 108 (1889), pp.\ 446--448

Halphen G-H [1] Sur les lois de Kepler, {\it Bulletin de la Soci\'et\'e Philomatique de Paris}, 7-1 (1878), pp.\ 89--91; {\it \OE uvres 2}, Paris: Gauthier-Villars, 1918, pp.\ 93--95

Hamilton WR [1] On a General Method in Dynamics; by which the Study of the
Motions of all free Systems of attracting or repelling Points is reduced to
the Search and Differentiation of one central Relation, or characteristic
Function, {\it Philosophical Transactions of the Royal Society}, 124 (1834), pp.\ 247--308; {\it Math.\ Papers},
vol.\ 2, pp.\ 103--161

Hamilton WR [2] Second Essay on a General Method in Dynamics, {\it Philosophical Transactions of the Royal Society}, 125 (1835),  pp.\ 95--144; {\it Math.\ Papers}, vol.\ 2, pp.\ 162--211

Killing W [1] Die Mechanik in den Nicht-Euklidischen Raumformen, {\it Journal f{\"u}r die reine und angewandte Mathematik}, 98 (1885), pp.\ 1--48

Lambert JH [1] {\it Insigniores Orbitae Cometarum Proprietates}, Augustae
Vindelicorum, Augsburg, 1761

 Levi-Civita T [1] Sur la r\'esolution qualitative du probl\`eme restreint des trois corps, {\it Verhandlungen des dritten Internationalen Mathematiker-Kongresses in Heidelberg vom 8. bis 13. august 1904}, Leipzig, 1905, pp.\ 402--408

MacLaurin C [1] {\it Treatise of Fluxions. In two books.} Ruddimans, Edinburgh, 1742,  \S 451 or \S 875

O'Neill B [1] {\it Semi-Riemannian Geometry, With Applications to Relativity}, Academic Press, 1983, p.\ 249

 Schr\"odinger E [1] A method of determining quantum-mechanical eigenvalues and eigenfunctions,  {\it Proc.\ Roy.\ Irish Acad.\ (Sect.\ A)} 46 (1940), pp.\ 9--16

Serret P  [1] {\it Th{\'e}orie nouvelle g{\'e}om{\'e}trique et m{\'e}canique des lignes {\`a} double courbure},  Mallet-Bachelier, Paris, 1859, p.\ 204

\end{document}